\documentclass[twocolumn,secnumarabic,amssymb, nobibnotes, aps, prd]{revtex4-1}

\usepackage{graphicx}
\usepackage[T1]{fontenc}
\usepackage[utf8]{inputenc}
\usepackage[colorlinks,linkcolor=blue,anchorcolor=blue,citecolor=blue]{hyperref}
\usepackage{indentfirst}
\usepackage{amsmath}
\begin{document}

\title{Exploring the photoproduction of $\rho$ and $\phi$ in hadronic heavy-ion collisions}

\author{Kaifeng Shen}
\author{Xin Wu}
\author{Zebo Tang}
\author{Wangmei Zha}
\email{first@ustc.edu.cn}
\affiliation{State Key Laboratory of Particle Detection and Electronics, University of Science and Technology of China, Hefei 230026,China}

\date{\today}

\begin{abstract}
\begin{@twocolumnfalse}

Significant enhancements of J/$\psi$ production have been observed by various experiments at RHIC and LHC for very low transverse momenta in peripheral heavy-ion collisions, which has ignited a surge of investigations into photon-induced processes in hadronic heavy-ion collisions (HHICs). Within this wave of research enthusiasm, the search for more photon induced products in HHICs becomes paramount.  In this paper, we perform the calculation of the $\rho$ and $\phi$ production resulting from photon-nucleus interactions in HHICs, which are crucial probes for studying the properties of Quark-Gluon Plasma (QGP) in HHICs. Our study reveals that, in comparison to hadronic production, the photon-induced production of $\rho$ and $\phi$ does not reach the same level of significance as that observed in J/$\psi$  production. Nevertheless, it remains substantial, especially in peripheral collisions, holding great promise for experimental verification in the imminent future.

\end{@twocolumnfalse}
\end{abstract}

\maketitle

\section{Introduction}
Understanding the intricacies of the Quark-Gluon Plasma (QGP) is crucial for unraveling the fundamental nature of matter under extreme conditions, such as those present in the early universe microseconds after the Big Bang. By utilizing the final-state particles resulting from relativistic heavy-ion collisions, notably those conducted at RHIC~\cite{STAR:2005gfr,Dong:2023nwv} and LHC~\cite{Niida:2021wut,Bala:2016hlf}, valuable insights into the properties and dynamics of the QGP can be gleaned, shedding light on the behavior of quarks and gluons in conditions of extreme temperature and energy density~\cite{Braun-Munzinger:2007edi}. Among the probes employed to investigate the QGP, vector mesons like $\rho$ and $\phi$ assume pivotal roles.  The measurements of $\rho$ (770) through leptonic decay is sensitive to the chiral symmetry restoration within the hot and dense quantum chromodynamics (QCD) environment, which establishes a connection to the origin of hadronic masses in the universe~\cite{Roberts:2007jh,Aoki:2013ldr}. Additionally, the study of the $\phi$ meson, bound state of strange quark and its anti-quark, proves especial insight for deciphering the QGP chemistry~\cite{Shor:1984ui, Mohanty:2009tz, Chen:2006ub, Nasim:2013fb}, as strangeness is not present in the valence content of the colliding nuclei and is light enough to be created copiously in the hot medium. 


\indent
In relativistic heavy-ion collisions, the production of vector mesons primarily arises from the intense strong interactions occurring within the nuclear overlap region; however, it extends beyond the realm of purely hadronic production. Another significant process, photon induced interaction, involving the emission of highly quasireal photons from one of the colliding nucleus~\cite{Bertulani:2005ru,Bottcher:1991nd,vonWeizsacker:1934nji}, could also generate vector mesons. These photons could fluctuate into $q\overline{q}$ pairs and subsequently interact with the other nucleus (coherent process) or the individual nucleons in the other nucleus (incoherent process) via Pomeron exchange, a color-neutral two-gluon state, and then emerge as vector mesons. In the coherent process, the final products consist of two intact nuclei, a vector meson with very low transverse momentum ($p_{\rm T}$ < 0.1 GeV/c) and nothing else. The exclusive photoproduction of a vector meson,  where both outgoing nuclei remain intact throughout the process, presents a distinctive opportunity to investigate the gluonic structure of nuclear matter. Notably, coherent interactions are conventionally believed to be only observable in Utra-Peripheral Collisions (UPCs), where the impact parameter (b) is larger than twice of the nuclear radius. In such collisions, the two colliding nuclei are expected to remain intact, satisfying the conditions necessary for coherent interactions. Therefore, the photoproduction of vector mesons was anticipated exclusively in UPCs, while their hadronic production was expected solely in hadronic heavy-ion collisions (HHICs)..
\newline
\indent
Is it possible that coherent photonucleus interacions also take place in HHICs, where the nuclei collide and disintegrate? The coherent photoproduction is a electromagnetic process that could occur at the periphery of the colliding nuclei where the influence of the hot medium is less pronounced. Consequently, the vector mesons produced via coherent photonucleus interactions may endure within the fireball and come to dominate in the very low $p_{\rm T}$ region. The experimental evidence was first revealed by the ALICE~\cite{adam2016measurement} and STAR~\cite{STAR:2019yox} collaborations, showing a significantly excess in the yield of J/$\psi$ at very low $p_{\rm T}$ (<0.3 GeV/c). Theoretical models~\cite{Zha:2017jch, PhysRevC.93.044912,Brandenburg:2022jgr} based on photoproduction processes have been employed to explain these excesses, since it is evident that these excesses can not be comprehended through known hot and cold medium effects in hadronic J/$\psi$ production. This research~\cite{Zha:2017jch} explores various photon and Pomeron coupling scenarios and addresses interference between J/$\psi$ photoproduction amplitudes from opposite directions. Besides the photonucleus interactions, STAR and ALICE also proved the coherent photon-photon interactions happened in HHICs by measuring the excesses in dilepton production at very low $p_{T}$~\cite{STAR:2018ldd,ALICE:2022hvk}, and the QED based calculations at leading-order~\cite{Zha:2018tlq,Brandenburg:2021lnj,Luo:2023syp} can describe these measurements in HHICs.
\newline
\indent
Can we observe additional photon-induced products in HHICs besides J/$\psi$, such as $\rho$ and $\phi$? If so, what are their contributions compared to their hadronic productions? Given that photoproduction has not been considered in previous experimental measurements, there is a question of whether it could impact the derived physical conclusions. Furthermore, can these photon-induced products also function as probes for studying the properties of the QGP? In this paper, we undertake calculations for the $p_{\rm T}$ spectrum of $\rho$ and $\phi$ photoproductions across various centrality bins at RHIC and LHC energies. Our objective is to determine the significance of photoproduction over the hadronic background, providing insights into the experimental challenges associated with verifying photon-produced $\rho$ and $\phi$ in HHICs.

\section{Methodology}
The cross section of the $\rho$ and $\phi$ mesons in the photonuclear process can be calculated by using the similar method in \cite{Zha:2020cst, PhysRevResearch.4.L042048}. For the experimental measurements, the production yields are usually performed in the transverse momentum space, especially for coherent photo-production case where the $p_{\rm T}$ of the photo-produced vector meson is constrained to be the level of the inverse of the colliding nucleus size~\cite{Andronic:2015wma}. In this case, the photo-produced $\rho$ and $\phi$ mesons will be dominated in the extremely low $p_{T}$ region which is about 60 MeV/$c$, while the yield of $\rho$ and $\phi$ mesons from hadronic processes decreases as $p_{\rm T}$ decreases in such low $p_{\rm T}$ region. In theoretical calculations, the amplitude distribution of the $\rho$ and $\phi$ mesons from the photonuclear process in two-dimensional transverse momentum representation can be calculated by performing a Fourier transformation to which in coordinate representation:
\begin{equation}
\setlength{\abovedisplayskip}{6pt}
\setlength{\belowdisplayskip}{6pt}
\begin{aligned}
\vec{A}(\vec{p}_{\bot})=&\frac{1}{2\pi}\int \, d^{2}x_{\bot}(\vec{A}_{1}(\vec{x}_{\bot})+\vec{A}_{2}(\vec{x}_{\bot})) \times\\
&e^{i\vec{p}_{\bot}\cdot \vec{x}_{\bot}},
\end{aligned}
\end{equation}
where $\vec{A}_{1}(\vec{x}_{\bot})$ and $\vec{A}_{2}(\vec{x}_{\bot})$ are the amplitudes distributions in the transverse plane from the two colliding nuclei, which can be obtained based on the corresponding $\gamma A \rightarrow VA$ scattering amplitude $\Gamma_{\gamma A \rightarrow VA}$ and the spatial photon flux in the outside the colliding nuclei~\cite{PhysRevC.97.044910}, shown as the following equation:

\begin{equation}
\setlength{\abovedisplayskip}{6pt}
\setlength{\belowdisplayskip}{6pt}
\begin{aligned}
A(\vec{x}_{\bot}) = \Gamma_{\gamma A \rightarrow VA}\sqrt{N(\omega_{\gamma},\vec{x}_{\bot})}.
\end{aligned}
\end{equation}

The scattering amplitude $\Gamma_{\gamma A \rightarrow VA}$ can be calculated by the Glauber model~\cite{Miller:2007ri} and vector meson dominance (VMD) approach~\cite{Bauer:1977iq}:
\begin{equation}
\setlength{\abovedisplayskip}{6pt}
\setlength{\belowdisplayskip}{6pt}
\begin{aligned}
\Gamma_{\gamma A \rightarrow VA}(\vec{x}_{\bot}) &= \frac{4\sqrt{\alpha}C}{f_{V}} \times \frac{f_{\gamma N \rightarrow VN}(0)}{f_{\gamma N \rightarrow VN}} \times \\
&2[1- {\rm exp}(-\frac{\sigma_{VN}}{2}T'(\vec{x}_{\bot}))],
\end{aligned}
\end{equation}
where $f_{V}$ is the $V$-photon coupling const and C is a correction factor for the non-diagonal coupling through higher mass vector mesons~\cite{Hufner:1997jg}, and $f_{\gamma N \rightarrow VN}(0)$ is the forward-scattering amplitude for $\gamma+N \rightarrow V+N$, which can be obtained from the well parameterized cross section of the forward-scattering $\frac{d\sigma_{\gamma N \rightarrow VN}}{dt}|_{t=0}$ as shown in Ref.~\cite{Klein:2016yzr}. The modified thickness function, $T'(\vec{x}_{\bot})$, is used to accounting for the coherent effect on $z$ direction:
\begin{equation}
\setlength{\abovedisplayskip}{6pt}
\setlength{\belowdisplayskip}{6pt}
\begin{aligned}
T'(\vec{x}_{\bot}) &= \int_{- \infty}^{+ \infty}\rho(\sqrt{\vec{x}_{\bot}^{2}+z^{2}})e^{iq_{L}z}dz,\\
q_{L} &= \frac{M_{\rho(\phi)}e^{y}}{2\gamma_{c}},
\end{aligned}
\end{equation}
where $q_{L}$ is the longitudinal momentum transfer required to produce a real vector meson, $\gamma_{c}$ is the Lorentz factor of the nucleus, $M_{\rho(\phi)}$ and $y$ are the mass and rapidity of the $\rho$ ($\phi$) meson. 

The distribution of photon flux generated by the colliding nuclei can be calculated based on the equivalent photon approximation (EPA):
\begin{equation}
\setlength{\abovedisplayskip}{6pt}
\setlength{\belowdisplayskip}{6pt}
\begin{aligned}
&\frac{d^{3}N_{\gamma}(\omega_{\gamma},\vec{x}_{\bot})}{d\omega_{\gamma}d\vec{x}_{\bot}}=\frac{4Z^{2}\alpha}{\omega_{\gamma}} \times  \\
&\left|\int \, \frac{d^{2}k_{\gamma\bot}}{(2\pi)^{2}}\vec{k}_{\gamma\bot}\frac{F_{\gamma}(\vec{k}_{\gamma})}{\left|\vec{k}_{\gamma}\right|^{2}}e^{i\vec{x}_{\bot}\cdot \vec{k}_{\gamma\bot}}\right|^{2} \\
&\vec{k}_{\gamma}=(\vec{k}_{\gamma\bot},\frac{\omega_{\gamma}}{\gamma_{c}}), \quad \omega_{\gamma}=\frac{1}{2}M_{\rho(\phi)}e^{\pm y},
\end{aligned}
\end{equation}
where Z is the nuclear charge number, $\alpha$ is the electromagnetic coupling constant, $\omega_\gamma$ is the energy of photon, $F_{\gamma}(\vec{k}_{\gamma})$ is the nuclear electromagnetic form factor and the $\vec{k}_{\gamma}$ is two-dimensional momentum vectors of the equivalent photons in the transverse momentum space. The form factor is obtained by the Fourier transformation of the charge density in the nucleus, which is described by the parameterized Woods-Saxon distribution:
\begin{equation}
\setlength{\abovedisplayskip}{6pt}
\setlength{\belowdisplayskip}{6pt}
\begin{aligned}
\rho_{A}(r) = \frac{\rho_{0}}{1+\rm exp \it [(r-R_{\rm WS})/\it d]},
\end{aligned}
\end{equation}
where $\rho_{0}$ is the normalization factor and the radius $R_{\rm WS}$ and skin depth $d$ are from electron-scattering experiments \cite{10.1063/1.2995107}. 

In addition, the incoherent photo-nucleus interactions also contribute to the production of $\rho$ and $\phi$ mesons in the very low $p_{\rm T}$ region, although the $p_{\rm T}$ of $\rho$ and $\phi$ mesons from incoherent photo-nucleus process is constrained to the order of the inverse of the nucleon size, which is about 300 MeV/$c$~\cite{Andronic:2015wma}. The cross section of the incoherent $\rho$ and $\phi$ mesons photoproduction, $\sigma_{\gamma+A \rightarrow V+A'}$, is scaled from the cross section $\gamma p \rightarrow Vp$ by the Glauber plus VMD approach, where $A'$ is the final state of the nucleus after the interaction, which contains the products of the nuclear disintegration. The cross section of incoherent photo-nucleus interaction can be calculated as:
\begin{equation}
\setlength{\abovedisplayskip}{6pt}
\setlength{\belowdisplayskip}{6pt}
\begin{aligned}
\sigma_{\gamma+A \rightarrow V+A'} &= \sigma_{\gamma+p \rightarrow V+p} \times \\
&\int d^{2}\vec{x}_{\bot} T(\vec{x}_{\bot})e^{-\frac{1}{2}\sigma^{in}_{VN}T(\vec{x}_{\bot})},\\
\sigma^{in}_{VN} &= \sigma_{VN}-\sigma^{2}_{VN}/(16\pi B_{V}),
\end{aligned}
\end{equation}
where $T(\vec{x}_{\bot})$ is the thickness function of the nucleus, $\sigma^{in}_{VN}$ is the inelastic vector meson-nucleon cross section and $B_{V}$ is the slope of the $t$ dependence of the $\gamma+p \rightarrow V+p$ scattering~\cite{Klein:2016yzr}.

\section{Results}

\begin{figure}[ht]
\centering

\includegraphics[width=0.9\columnwidth,clip]{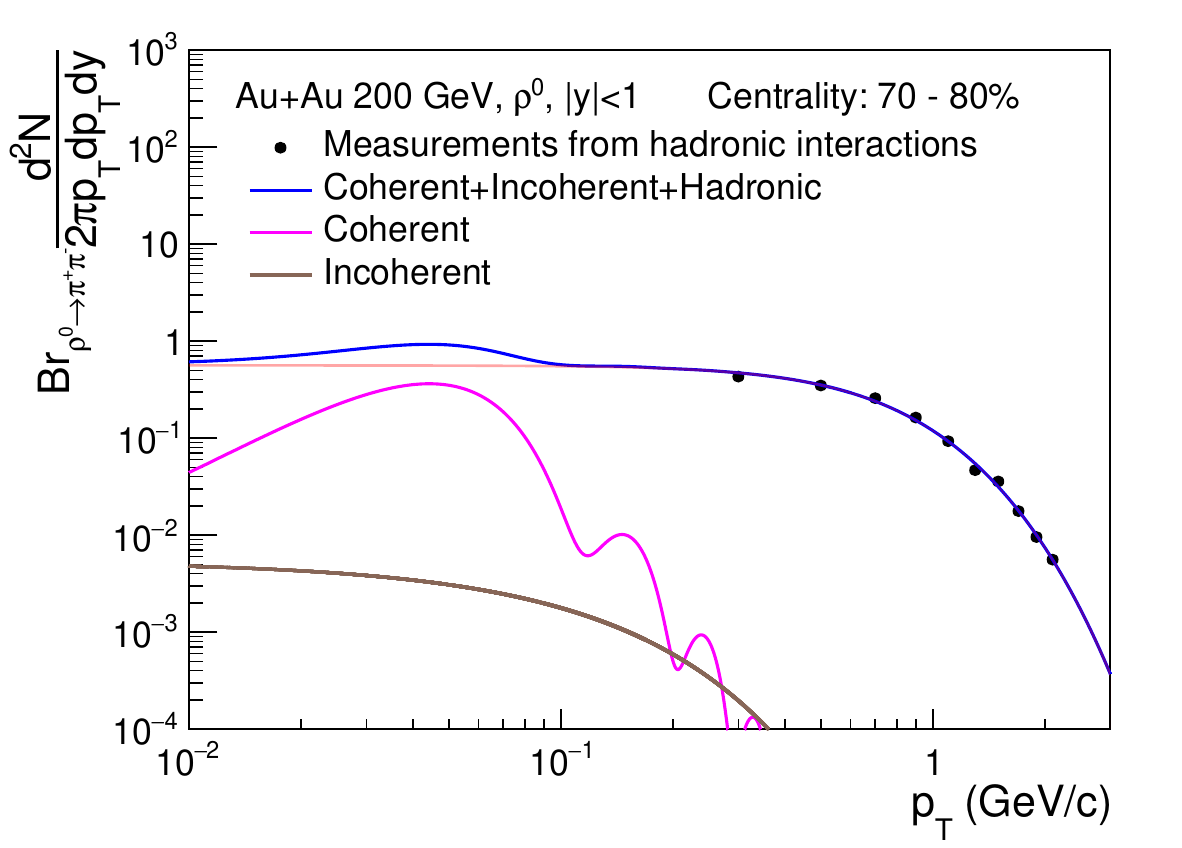}

\includegraphics[width=0.9\columnwidth,clip]{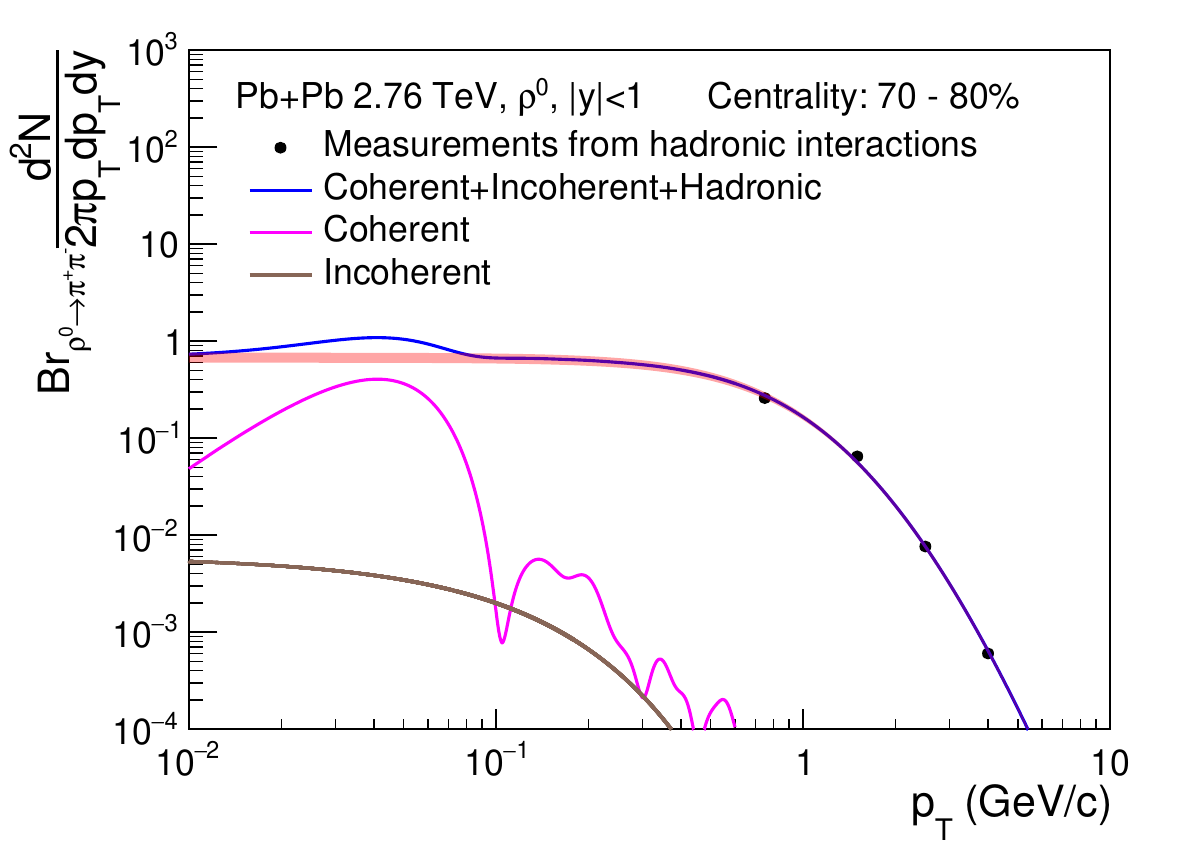}

\caption{The calculated invariant yield of $\rho$ meson resulting from photoproduction as a function of $p_{\rm T}$ in 70-80$\%$ Au+Au collisions at $\sqrt{s_\mathrm{NN}} =$ 200 GeV (Top) and in Pb+Pb collisions at $\sqrt{s_\mathrm{NN}} =$ 2.76 TeV (Bottom). The magenta solid lines represent the contribution originating from the coherent photon-nucleus process, while the brown solid lines denote the production derived from the incoherent photon-nucleus process. The blue solid lines stand for the total yield of $\rho$ mesons, encompassing contributions from both photoproduction and hadronic production.}
\label{fig-1}       
\end{figure}

Figure~\ref{fig-1} illustrates the calculated invariant yield distributions of $\rho$ mesons, $Br._{\rho\rightarrow\pi^{+}\pi^{-}}\frac{d^{2}N}{2\pi dp_{T}dy}$, arising from photoproduction in 70-80$\%$ Au+Au collisions at $\sqrt{s_\mathrm{NN}} =$ 200 GeV and Pb+Pb collisions at $\sqrt{s_\mathrm{NN}} =$ 2.76 TeV. The magenta solid lines depict the distribution of photo-produced $\rho$ mesons resulting from coherent interactions, while the brown solid lines represent those resulting from incoherent interactions. The blue solid lines indicate the total yields of $\rho$ mesons, encompassing contributions from both photoproduction and hadronic interactions. As observed in Fig.~\ref{fig-1}, the $p_{\rm T}$ distribution of coherent photo-produced $\rho$ mesons is highly concentrated in the extremely low range ($p_T < 0.1$ GeV/$c$). The decreasing trend as $p_{\rm T}$ approaches zero is a consequence of interference effects~\cite{Zha:2020cst}. The peak structures originating from the coherent photon-nucleus process are attributed to the combined effects of interference and diffraction, and noticeable differences between Au+Au and Pb+Pb collisions are evident, stemming from the distinct size and shape of the two nuclei. For the incoherent photoproduction case, the $p_{\rm T}$ distribution extends to a relatively higher range, as revealed by our calculations. Experimental measurements of $\rho$ meson yields resulting from hadronic interactions in Au+Au at $\sqrt{s_\mathrm{NN}} =$ 200 GeV and in Pb+Pb collisions at $\sqrt{s_\mathrm{NN}} =$ 2.76 TeV are represented by the black points~\cite{ALICE:2018qdv,STAR:2003vqj}.


\begin{figure}[ht]
\centering

\includegraphics[width=0.9\columnwidth,clip]{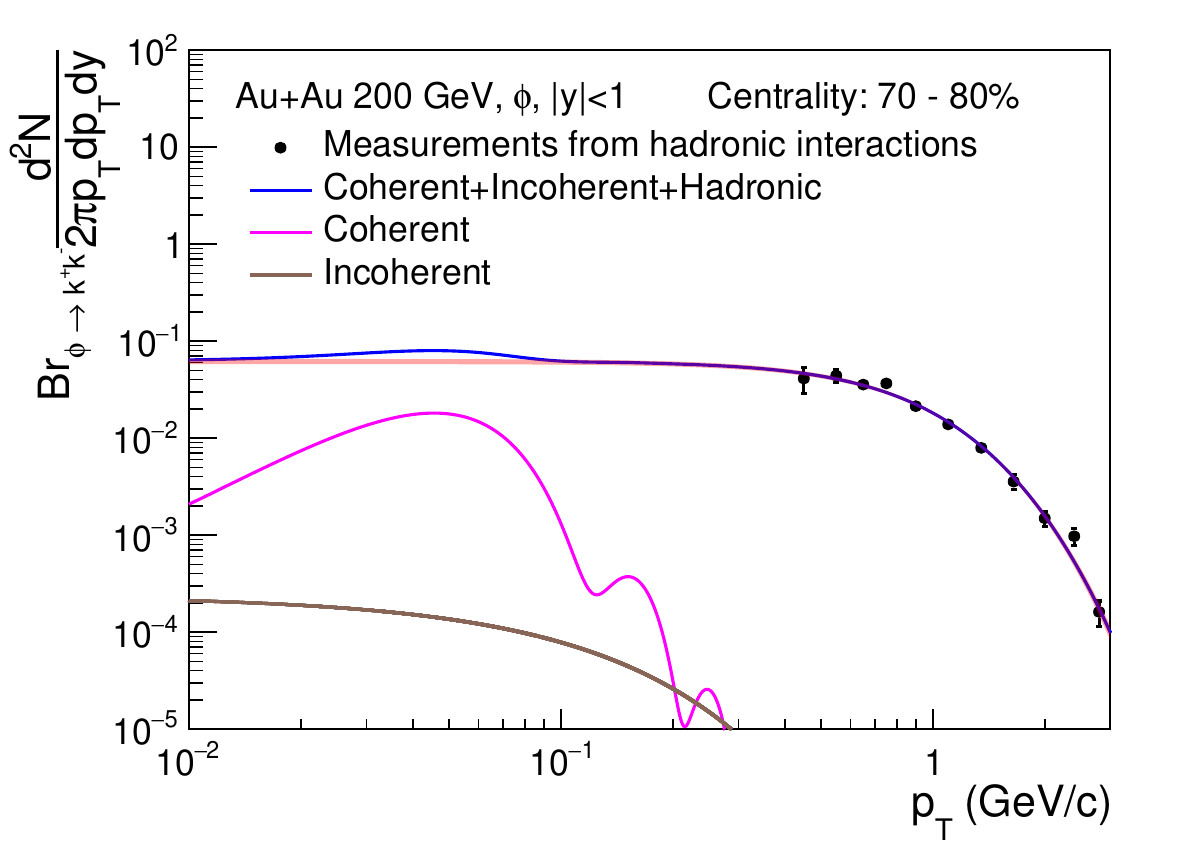}

\includegraphics[width=0.9\columnwidth,clip]{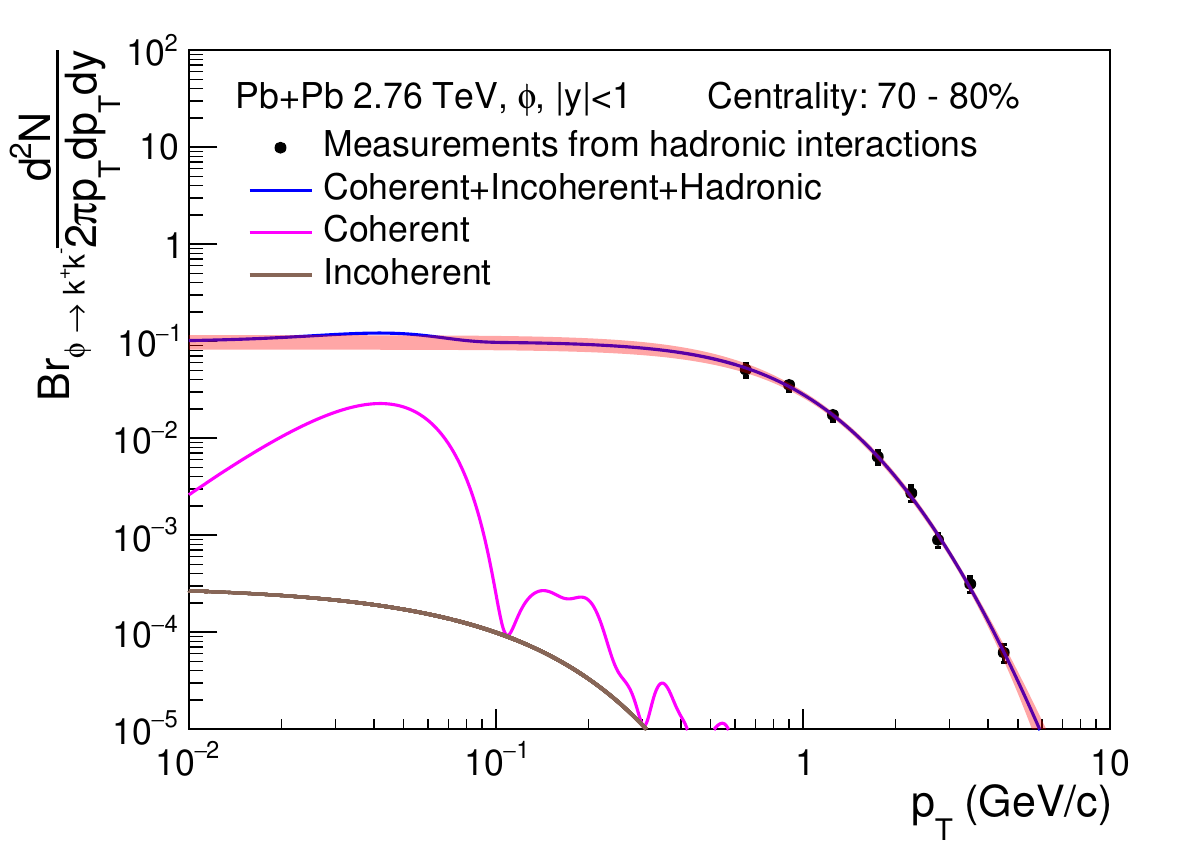}

\caption{The calculated invariant yield of photon produced $\phi$ meson as a function of $p_{T}$ in 70-80$\%$ Au+Au collisions at $\sqrt{s_\mathrm{NN}} =$ 200 GeV (Top) and Pb+Pb collisions at $\sqrt{s_\mathrm{NN}} =$ 2.76 TeV (Bottom). The magenta solid lines represent the invariant yield originating from the coherent process, while the brown solid lines present similar results derived from incoherent interactions. The blue solid lines denote the total yield of $\phi$ mesons, encompassing contributions from both coherent and incoherent processes.}
\label{fig-2}       
\end{figure}

To estimate the yield of $\rho$ mesons from hadronic production at extremely low transverse momentum ($p_{\rm T}$), the experimental data in Fig. \ref{fig-1} is subjected to fitting using the following formula:
\begin{equation}
\setlength{\abovedisplayskip}{6pt}
\setlength{\belowdisplayskip}{6pt}
\begin{aligned}
&\frac{1}{2\pi p_{T}}\frac{d^{2}N}{dydp_{T}}=A\cdot e^{[-B(\sqrt{p^{2}_{T}+m^{2}_{V}}-m_{V})]}
\end{aligned}
\end{equation}
where A and B are free parameters and $m_{V}$ is the pole mass of the vector meson.  The fitted results are presented as the red band in Fig. \ref{fig-1}, where the band conveys the uncertainties associated with the fitting process. Through these extrapolated results, the yield of photo-produced $\rho$ mesons is found to be comparable to that from hadronic interactions at very low $p_{\rm T}$. Notably, the $\rho$ meson yield from photoproduction closely aligns with that from hadronic production in 70-80$\%$ Au+Au collisions at RHIC and Pb+Pb collisions at LHC. The comparable invariant yield results for $\phi$ mesons are exhibited in Fig. \ref{fig-2}, revealing that the yield fraction of $\phi$ mesons from photoproduction is lower than that of $\rho$ mesons in both collision systems.

\begin{figure}[ht]
\centering
\includegraphics[width=0.9\columnwidth,clip]{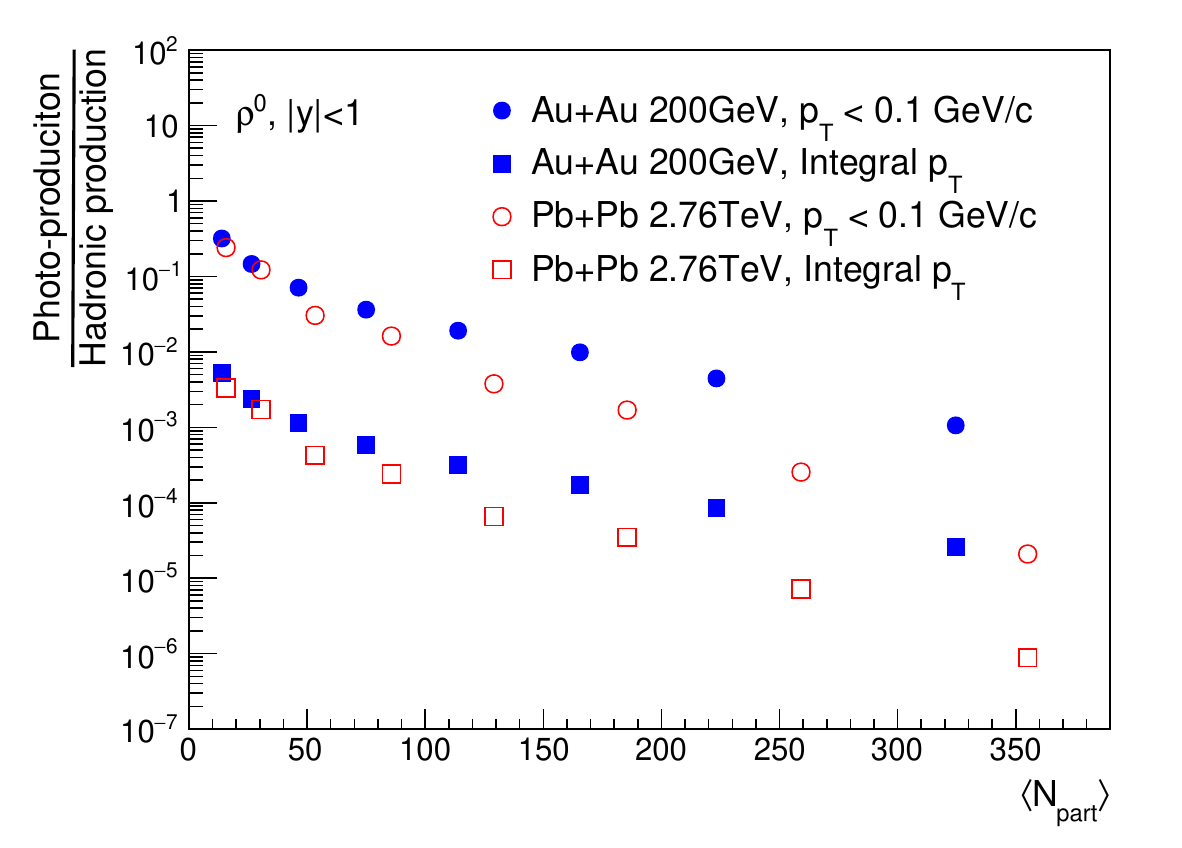}
\caption{The ratio of photo-produced $\rho^{0}$ mesons to those originating from hadronic interactions  as a function of $\langle N_{\rm{part}} \rangle$, for $p_{\rm T} <$ 0.1 GeV/$c$ and integrated $p_{\rm T}$ in Au+Au collisions at $\sqrt{s_\mathrm{NN}} =$ 200 GeV (depicted by blue markers) and in Pb+Pb collisions at $\sqrt{s_\mathrm{NN}} =$ 2.76 TeV (depicted by red markers).}
\label{fig-3}
\end{figure}

\begin{figure}[ht]
\centering
\includegraphics[width=0.9\columnwidth,clip]{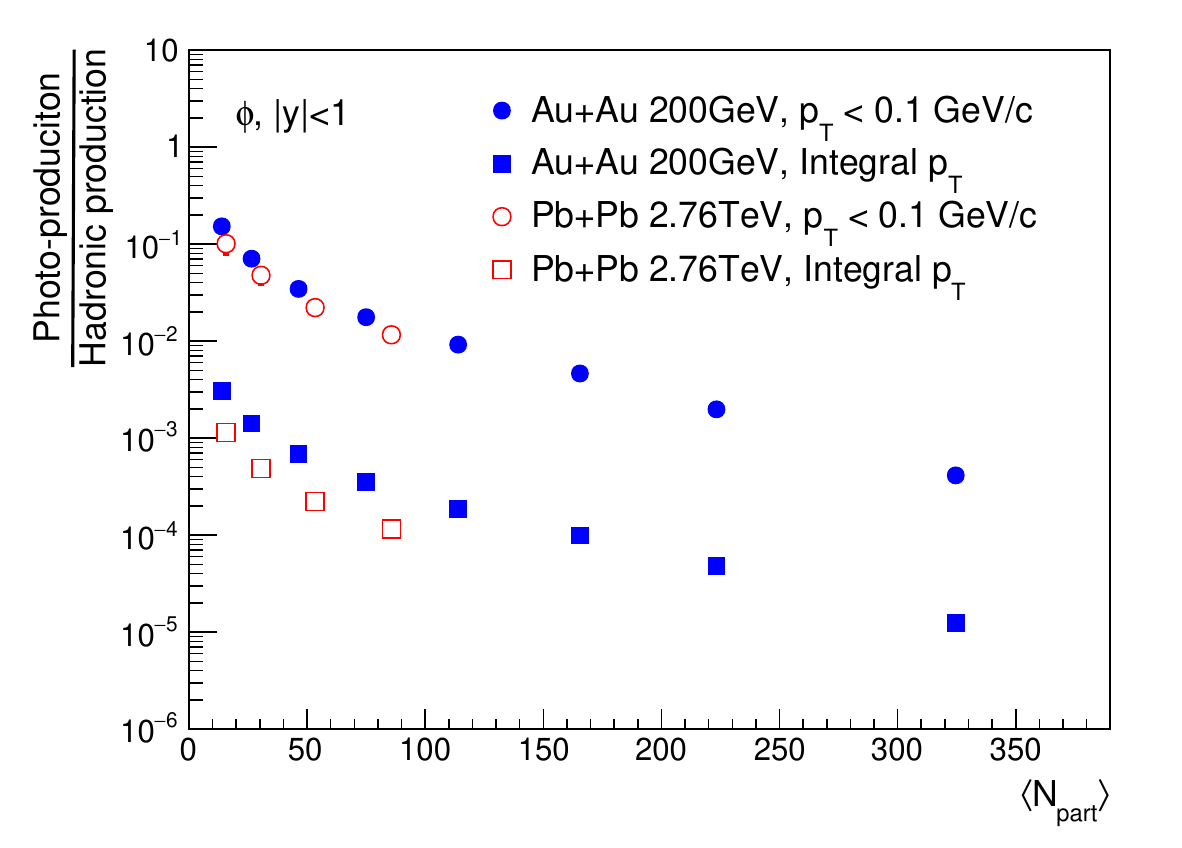}
\caption{The ratio of photo-produced $\phi$ mesons to those originating from hadronic interactions  as a function of $\langle N_{\rm{part}} \rangle$, for $p_{\rm T} <$ 0.1 GeV/$c$ and integrated $p_{\rm T}$ in Au+Au collisions at $\sqrt{s_\mathrm{NN}} =$ 200 GeV (depicted by blue markers) and in Pb+Pb collisions at $\sqrt{s_\mathrm{NN}} =$ 2.76 TeV (depicted by red markers).}
\label{fig-4}
\end{figure}

Figure~\ref{fig-3} illustrates the ratio of photo-produced $\rho$ mesons to those originating from hadronic production, presenting a dependence on $\langle N_{\rm{part}} \rangle$, at very low $p_{\rm T}$ (< 0.1 GeV/$c$) and integrated $p_{\rm T}$ in Au+Au collisions at $\sqrt{s_\mathrm{NN}} =$ 200 GeV and Pb+Pb collisions at $\sqrt{s_\mathrm{NN}} =$ 2.76 TeV. Different colored data points distinguish between various collision systems. The observed trend reveals an increase in the ratio as $\langle N_{\rm{part}} \rangle$ decreases, suggesting a more pronounced photo-production of $\rho$ mesons in peripheral nucleus-nucleus collisions. In central collisions, contributions from photo-production appear negligible; however, in peripheral collisions, photo-induced production accounts for approximately 30$\%$ of $\rho$ production originating from hadronic interactions at very low $p_{\rm T}$ (< 0.1 GeV/$c$). The integrated yield of $\rho$ meson from photoproduction is much smaller than those from hadronic interactions in all centrality classes, which suggests that the additional contribution from photoproduction does not significantly impact the derived physical conclusions from previous experimental measurements. Moving on to the ratios of photo-produced $\phi$ mesons to those from hadronic interactions, depicted as a function of $\langle N_{\rm{part}} \rangle$ at $p_{\rm T} <$ 0.1 GeV/$c$ and integrated $p_{\rm T}$ in Fig.~\ref{fig-4}, we observe a similar trend to that of $\rho$ mesons. However, the proportion of photo-production for $\phi$ mesons is relatively lower compared to that of $\rho$. In the most peripheral Au+Au collisions at $\sqrt{s_\mathrm{NN}} =$ 200 GeV, the photo-induced production of $\phi$ mesons reaches up to 15$\%$ of the $\phi$ yield originating from hadronic interactions for very low $p_{\rm T}$ (< 0.1 GeV/$c$). While $\rho$ and $\phi$ photoproduction is not as pronounced as that of J/$\psi$ at very low $p_{\rm T}$ in peripheral collisions, it remains comparable to that from hadronic interactions on the order of magnitude. This suggests promising avenues for experimental verification in the near future, necessitating higher statistical requirements. However, it is crucial to note that this study represents an initial step, and more advanced models are required to explore potential hot medium effects on $\rho$ and $\phi$ photoproduction.
\section{Summary}
In summary, this study delves into the realm of $\rho$ and $\phi$ photoproduction within in hadronic high-energy heavy-ion collisions at both RHIC and LHC energies. Our calculations compellingly underscore that the yield of $\rho$ and $\phi$ at very low $p_{\rm T}$ has significant contribution from coherent photon-nucleus process. Methodically considering coherent and incoherent photo-produced $\rho$ and $\phi$ mesons and juxtaposing them with extrapolated hadronic contributions yields are valuable insights. Compared to its hadronic counterpart, the integrated yield from photoproduction registers as negligible, which suggests that the additional contribution from photoproduction does not wield a significant impact on the established physical conclusions drawn from preceding experimental measurements. However, at $p_{\rm T} < 0.1$ GeV/$c$ in peripheral collisions, this yield remains comparable to that from hadronic interactions. This holds promise for experimental verification of $\rho$ and $\phi$ photoproduction in HHICs in the imminent future, yet it demands augmented statistical requirements. A pivotal query arises: Can these photon-induced products serve as probes for probing the properties of the QGP? This question sparks curiosity and beckons further investigation. It is paramount to acknowledge that this study constitutes an inaugural exploration, and the pursuit of more advanced models becomes imperative to scrutinize potential hot medium effects on $\rho$ and $\phi$ photoproduction.

\section{Acknowledgments}
This work is supported in part by the National Key Research and Development Program of China under Contract No. 2022YFA1604900 and the National Natural Science Foundation of China (NSFC) under Contract No. 12175223 and 12005220. W. Zha is supported by Anhui Provincial Natural Science Foundation No. 2208085J23 and Youth Innovation Promotion Association of Chinese Academy of Science.

\bibliographystyle{unsrt}
\bibliography{reference.bib}

\end{document}